\documentclass[final,5p,twocolumn,sort&compress,letterpaper]{elsarticle}
\usepackage{jeremy_elsevier}

\begin{document}

% frontmatter
\begin{frontmatter}

	%--------------------------------------------------
  	% TITLE
  	%--------------------------------------------------
	\title{
	    \texorpdfstring{Identification of weakly to strongly-turbulent three-wave processes\\in a micro-scale system}
        {Identification of weakly- to strongly-turbulent three-wave processes in a micro-scale system}
	    }

	%--------------------------------------------------
	% AUTHORS AND AFFILIATIONS
	%--------------------------------------------------
    \author[]{J.~Orosco}
    \author[]{W.~Connacher}
    \author[]{J.~Friend\corref{cor1}}
    \cortext[cor1]{Corresponding author, jfriend@ucsd.edu}
    \address{
        Medically Advanced Devices Laboratory, Center for Medical Devices\\
        Department of Mechanical and Aerospace Engineering
        University of California San Diego,
        La Jolla, CA 92093-0411 USA
    }

	%--------------------------------------------------
	% ABSTRACT
	%--------------------------------------------------
	\begin{abstract}

		We find capillary wave turbulence (WT) to span multiple dynamical regimes and geometries---from weakly to strongly nonlinear WT (SWT) and from shallow to deep domains---all within a $40\,\upmu$L volume millifluidic system. This study is made viable with recent advances in ultra-high-speed digital holographic microscopy, providing 10-$\upmu$s time and 10-nm spatial resolutions for images across the entire field of view, and encompassing a complete wave system at speeds sufficient to capture the salient wave phenomena. We provide a set of tractable parameters that identify the four fundamental WT regimes present in this simple system. A proposed nonlinearity measure permits comparative analysis while varying input conditions. This work augments current understanding of WT regimes and behaviors, and directly applies to many fields beyond fluid mechanics. For example, SWT appears upon the fluid interface at powers less than required for atomization, indicating that further study of SWT is needed to properly understand ultrasound-driven fuel spray atomization and drug and agricultural nebulization. %The curated data ($300\,$GB) obtained for this study are freely provided for download~\cite{orosco_data_2022}.
\end{abstract}

%--------------------------------------------------
% KEYWORDS
%--------------------------------------------------
\end{frontmatter}

%--------------------------------------------------
% INTRODUCTION
%--------------------------------------------------
\section{Introduction}
Applications from ink jet printing to fuel combustion depend on rapid, monodisperse droplet production~\cite{friend_microscale_2011,connacher_droplet_2020}. Many depend on atomization to produce micron-sized droplets from small fluid volumes~\cite{connacher_micro/nano_2018}. Leveraging extraordinarily large accelerations, high-frequency ultrasound (HFUS) at 1~MHz and beyond extends the utility of ultrasound-driven atomization to a broader range of fluid parameters \cite{Collignon:2018cs}. However, the phenomena is poorly understood and even basic predictions---such as the atomized droplet diameter and ejection rate---are still impossible to provide. In particular, classic interpretations of the ultrasound-driven atomization phenomena \cite{lang62,Rajan:2001vt} that rely on sophisticated modeling \cite{faraday_xvii_1831,Benjamin:1954ue,Miles:1992p1493,Kumar:1996tz} yet do not provide accurate estimates at frequencies beyond 100~kHz. More recent, ad-hoc approaches produce interesting results but do little better in predicting the atomization phenomena \cite{Qi:2008rp,Collignon:2018cs}. 

One of the key problems is the lack of understanding the nature of the waves present on the fluid interface, especially with excitation that drives motion beyond the typical linear coupling seen in droplet vibration experiments~\cite{Song:2022uv}. Consider the HFUS oscillation of a simple fluid parcel that produces atomization from its surface. Faraday waves~\cite{miles1984nfr,miles1990parametrically} have long been hypothesized to be responsible for HFUS atomization \cite{lang62,Tsai:2012uj,Yuan:2022vs}. These waves oscillate at one half the driving frequency and exhibit superharmonics at frequency-doubled intervals~\cite{xia_modulation_2010}. The waves also generate complex patterns that have drawn significant interest over the years~\cite{Bosch:1994vv}, mainly due to their connection to chaotic phenomena and the formation of elegant patterns \cite{cross1993pattern}. 

However, there is a fundamental problem with the presumption that Faraday waves are responsible for HFUS atomization: the assumption that the excited capillary waves' frequency is of the same order as the excitation frequency does not hold. In HFUS-driven capillary waves, an isolated, Lorentzian response peak appears and matches the driving frequency: a linear response. There is no response at one-half the excitation frequency as one would expect with Faraday waves. Nor are there any other response peaks at rational fractions of the excitation frequency that one might further expect from parametric coupling, as an extension of Faraday waves \cite{PhysRevLett.47.1133,miles1990parametrically}. Instead, there is a procession of Lorentzian response peaks---a nearly linear modal superposition response with modes that correspond to the Rayleigh-Lamb equation for oscillation of the parent droplet---at a much lower frequency range than the original HFUS excitation~\cite{xia_modulation_2010,blamey_microscale_2013}. 

Moreover, when driven by vibration at an amplitude beyond a threshold defined by the fluid and device characteristics, the nearly linear modal superposition response at low frequencies gives way to a continuously distributed energy cascade~\cite{xia_modulation_2010,blamey_microscale_2013}, apparently an example of capillary wave turbulence \cite{zakharov_kolmogorov_1992}. The hallmark of this turbulent cascade is a well-defined, monotonically decreasing, non-integer linear slope in the log-log power spectral density (PSD). Crucially, then, the phenomenon of HFUS-driven capillary waves and atomization appears to depend upon wave turbulence.
% FIGURE ONE
% ONE COLUMN FIGURE
% experimental setup
\begin{figure}
    \begin{center}
        \includegraphics[width=\columnwidth]{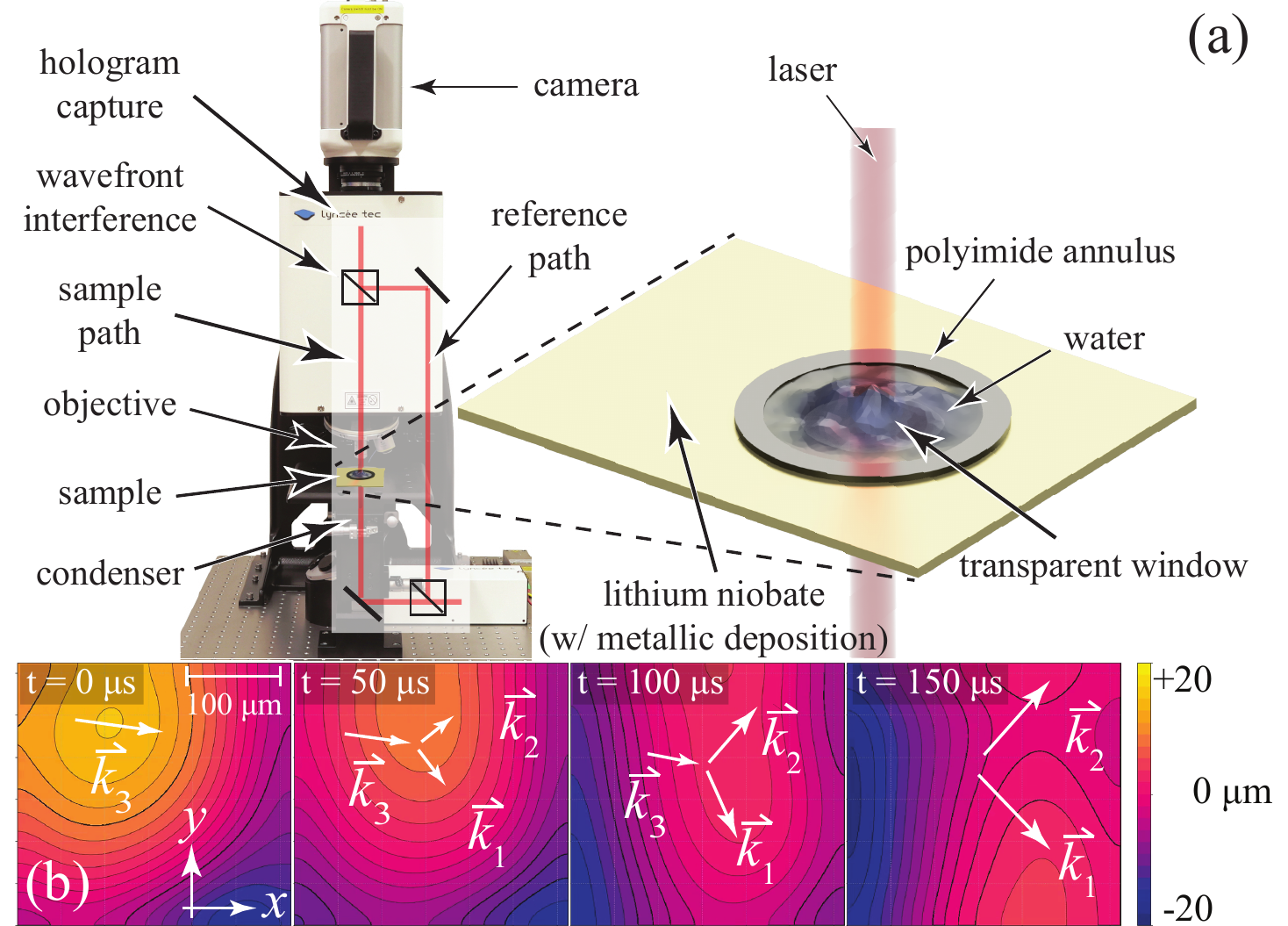}
        \caption{High-speed digital holographic microscopy of turbulent microscale capillary waves. (a): An interferometer using a $666\,$nm laser expanded to occupy the entire field of view of the optics produces phase and intensity patterns that encode 3D surface data recorded at $115.2\,$kfps. A $25\,\times\,20\,\times\,0.5\,$mm piezoelectric transducer was fabricated using single-crystal, transparent lithium niobate with a $6.4\,$mm window in the electrodes facilitating laser passage. It was driven at its resonant frequency of $7\,$MHz, causing vibration below a $725\,\upmu$m deep fluid volume contained within a $9.5\,$mm diameter annulus. (b) Four frames from a typical holographic data set of interfacial dynamics depicts a typical three-wave interaction.}
        \label{fig:f01_exp_setup}
    \end{center}
\end{figure}

Wave turbulence (WT) is of great interest across many contexts, from brain activity patterns~\cite{sheremet_wave_2019} to optical wave propagation within nonlinear media~\cite{picozzi_optical_2014}. In pioneering work, Zakharov and Filonenko~\cite{zakharov_weak_1967} derived a kinetic equation governing capillary WT cascades in unbounded basins by expanding a Hamiltonian representing the interfacial dynamics to third order in the wave amplitude. This equation governs conservative dynamics from the ``inertial'' range down to the beginning of the viscous ``dissipation'' range. They show that the stationary collision integral identity is satisfied by $n(k)\propto k^{-\gamma}$, where $n$ is proportional to the square of the wave amplitude at wavenumber $k$. The system in Ref.~\cite{zakharov_weak_1967} has $\gamma=17/4$. Due to the governing kinetic equation, this is denoted ``kinetic'' WT (KWT). Strong WT (SWT), by contrast, is WT that violates the weak nonlinearity assumption defined in terms of the spatial surface gradient. This gradient corresponds to the steepness of the capillary waves. It is important to note that these results are for unbounded capillary wave interfaces that have infinite spatial extent.

Within finite geometries, WT behaviors deviate from the above ideal. Wave modes form a countably infinite set and the associated WT is designated ``discrete'' WT (DWT). Kartashova~\cite{kartashova_nonlinear_2010} showed that exact three-wave equations satisfied in capillary KWT become Diophantine in DWT and are equivalent to $x^3+y^3=z^3$, which has no solution (\emph{see} Fermat's Last Theorem \cite{Kleiner:2000uk}). In the absence of exact resonances, waves build to finite amplitude and satisfy nonlinear dispersion relations~\cite{debnath_nonlinear_1994,pushkarev_turbulence_2000} leading to quasi-resonant three-wave relations~\cite{connaughton_discreteness_2001}:
%
% quasi-resonant three-wave equations
\begin{subequations}
    \begin{align}
        |\omega_{k_1}\pm\omega_{k_2} - \omega_{k_3}| &\leq \delta_{\textsc{NRB}},\\
        {\bf{k}}_1 \pm {\bf{k}}_2 - {\bf{k}}_3 &= 0,
    \end{align}
\end{subequations}
where $\omega_{k_n}\triangleq\omega({\bf{k}}_n)$. The nonlinear resonance broadening (NRB) is represented here by a parameter $\delta_{\textsc{NRB}}$ in the finite-amplitude dispersion relation. Upon unbounded media, $\delta_{\textsc{NRB}}=0$ and exact resonances appear as arguments to Dirac combs within a collision integral. In finite media, $\delta_{\textsc{NRB}}>0$. When the capillary wave amplitude is sufficiently large, the equations are satisfied by groupings of quasiresonances about exact modes. Modal broadening facilitates additional wave interactions and turbulent breakdown.

Small $\delta_{\textsc{NRB}}$ values generate arrested cascades called ``frozen'' turbulence. Pushkarev and Zakharov~\cite{pushkarev_turbulence_2000} show that this leads to layered ``wedding cake'' distributions in wavenumber space~\cite{pushkarev_turbulence_2000,connaughton_discreteness_2001}. Connaughton \etal~\cite{connaughton_discreteness_2001} predict a critical NRB value beyond which cascades  indefinitely advance. Cascade advancement from larger forcing---called ``sandpiling''---has been described by Nazarenko~\cite{nazarenko_sandpile_2006}.

In real systems, DWT and KWT coexist~\cite{lvov_discrete_2010,kartashova_model_2006,pushkarev_turbulence_2000,zakharov_mesoscopic_2005}. The dominance of either mechanism depends on the wavenumber and the forcing amplitude. A system exhibiting these mechanisms may be classified as follows:
%
% list of wave turbulence regimes
\begin{enumerate*}[label=(\roman*)]
    \item\label{it:d_regime} at low powers and/or small wavenumbers, where DWT dominates;
    \item\label{it:k_regime} at high powers and/or large wave numbers, where KWT dominates; and
    \item\label{it:m_regime} at intermediate powers and/or intermediate wavenumbers, where a combination of DWT and KWT contribute.
\end{enumerate*}
Here, we designate condition \ref{it:m_regime} as \emph{intermediate WT} (IWT), though it has also been called ``mesoscopic'' WT~\cite{zakharov_mesoscopic_2005}.

The majority of liquid WT experiments are devoted to gravity waves; surface tension-dominant systems receive less attention. Work that \textit{has} considered capillary waves generally takes place in deep water environments, often in the presence of gravity waves, and focuses on regimes where kinetic theory is approximately satisfied~\cite{brazhnikov_observation_2002,falcon_observation_2007-1,herbert_observation_2010,kharbedia_moulding_2021,snouck_turbulent_2009,xia_modulation_2010}. Explicit consideration of finite-domain effects has remained mostly theoretical, with Pan and Yue~\cite{pan_understanding_2017} recently providing a framework to account for discreteness by extending earlier work in the kinetic theory of deep-water capillary waves~\cite{pan_direct_2014,pan_decaying_2015}.

Here, we study a surface tension-dominated system where spatial scales are much smaller than the capillary length, $k_*^{-1}=\sqrt{\sigma/\rho\,g}$. The device geometry used in our experiments approximates the dimensions of moderately shallow, surface-wetting ``puddles'' that form on prototype portable ultrasonic nebulizers developed in our lab~\cite{Collignon:2018cs,huang_practical_2021} (\emph{see} [supplemental materials]). We can measure these dynamics using a custom, ultra high-speed (UHS) digital holographic microscope (DHM) that provides 10~microsecond-3~nanometer displacement resolutions over the entire $300\times 300\,\upmu$m$^2$ field of view with a lateral resolution of 1.2~$\upmu$m. Without this capability, important portions of the cascade would remain inaccessible. Due to the range of scales, geometries, and wavenumbers, the system passes through several quantitatively-significant physical regimes captured using this approach:
%
% list of transitions
\begin{enumerate*}[label=(\roman*)]
    \item\label{it:depth_trans} fluid depth: shallow, intermediate, and deep;
    \item\label{it:wt_trans} dominant WT dynamics: discrete, intermediate, and kinetic; and
    \item\label{it:nonlinear_trans} nonlinearity level: weak and strong.
\end{enumerate*}
We later derive dimensionless quantities from discrete and kinetic theory to analyze and classify the WT dynamics across each of these regimes.

%--------------------------------------------------
% EXPERIMENT
%--------------------------------------------------
\section{Experiment}

The experimental configuration is outlined in Fig.~\ref{fig:f01_exp_setup}. The device is a $25\,$mm$\,\times\,20\,$mm$\,\times\,0.5\,$mm single-crystal, transparent lithium niobate piezoelectric transducer with electrodes deposited on each face leaving a $6.35\,$mm diameter window. In order to repeatably produce a fluid sample upon this substrate, a $60\,\upmu$m thick polyimide annulus with inner diameter $9.5\,$mm was affixed to the top face, encircling the window. Deionized water ($40\,\upmu$L) was pipetted into the annulus such that a thin circular lens with maximum central depth $\approx725\,\upmu$m completely filled the polyimide annulus, placing the contact line at the top edge of the annulus. A sinusoidal voltage signal was applied at $7.001\,$MHz, driving the fundamental thickness-mode resonance in the transducer \cite{Vasan:2020fu}. Greater detail regarding materials, fabrication, and characterization is available in [supplemental materials].

The central portion of the air-water interface was then imaged using an UHS camera (Photron, SA-Z) coupled to a DHM (Lynce\'e~Tec~SA, Lausanne, Switzerland) with custom optics designed by Lynce\'e~Tec~SA for these experiments. High-intensity coherent light from a 666~nm laser is equally split between measurement light passing through the sample and reference light passed unhindered around the sample. Light passing through the sample medium encounters a phase delay with respect to the reference. Using the sample's refractive index, the phase delay may be associated with a displacement up to $2\,\pi\,\lambda$ where $\lambda$ is the light source's wavelength. Phase jumps exist at integer multiples of $2\,\pi\,\lambda$; left unaccounted for, they create ambiguity in predictions of the displacement of the sample's medium. However, if the height gradually changes with respect to the viewing plane, phase jumps that occur can be \emph{unwrapped}---accounted for---to produce high-fidelity surface displacement measurements well beyond the $2\,\pi\,\lambda$ limit. We obtained surface holograms covering a $300\,\upmu$m$\,\times\,300\,\upmu$m$^2$ square central region of the oscillating fluid interface. Holograms were recorded at $115.2\,$kfps with a $10\,$nm displacement resolution along the light propagation direction, and a $1.2\,\upmu$m lateral image plane resolution. The curated data ($300\,$GB) obtained for this study are freely provided for download~\cite{orosco_data_2022}.

%--------------------------------------------------
% WAVE TURBULENCE REGIME CLASSIFICATION
%--------------------------------------------------
\section{Classifying wave turbulence regimes}
We begin with an approach similar to Zakharov~\cite{zakharov_statistical_1999} and L'vov and Nazarenko~\cite{lvov_discrete_2010}. They formed an intuitive comparison of the nonlinear resonance broadening (NRB), $\delta_{\textsc{NRB}}$, to the spacing, $\delta_{k}$, in the eigenmode grid that is imposed by the finite geometry. DWT dominates the wavemode behavior when $\delta_{\textsc{NRB}}\ll\delta_{k}$. KWT dominates the wavemode behavior when $\delta_{\textsc{NRB}}\gg\delta_{k}$. Otherwise, the characteristics of both types may be observed, leading to IWT.

For the purposes of inter-spectrum classification and comparison, we define an expression for bicoherence, a bulk nonlinearity metric based on a weighted average over the three-wave correlated dynamical measure. We begin with a brief review of the relevant details of weak wave turbulence theory as the basis for the definitions that follow.

%--------------------------------------------------
% BACKGROUND
%--------------------------------------------------
\subsection{Background}
The eigenmode grid spacing at a particular wavenumber $\kappa$ may be written in terms of the smallest possible wave number, $k_m=\pi/L$, as
%
% grid-spacing determination
\begin{align}\label{eq:grid_spacing}
    \begin{split}
        \delta_{k}/k_m&\approx d\omega/dk,\\
        \Rightarrow\delta_{k}&\approx k_m\frac{d\omega}{dk}\biggr|_{\kappa},
    \end{split}
\end{align}
for a container of width $L$ and where we have assumed homogeneous Dirichlet boundaries. The form of the dispersion relation---$\omega_k=\omega(\bm{k})$, with $k=\ltnorm{\bm{k}}$ and where $\ltnorm{(\,\cdot\,)}$ is the $\ell^2$-norm of $(\,\cdot\,)$---depends on the particular system. In this work, we encounter both shallow and deep water capillary waves. We distinguish between these using an inverse capillary length, $k_*=\sqrt{\rho\,g/\sigma}$, and the fluid depth, $h$. If $k_*\ll k\ll 1/h$, the conditions are shallow. If $k\gg 1/h$, the conditions are deep~\cite{zakharov_kolmogorov_1992}.
            
The accepted approach to modeling a conservative dynamical regime of wave turbulence is to consider the Hamiltonian, $\mathcal{H}$, of the interfacial dynamics in canonical Fourier coordinates~\cite{zakharov_kolmogorov_1992}:
%
% canonical fourier hamiltonian equation
\begin{align}\label{eq:hamiltonian_equation}
    i\,\frac{da_{\bm{k}}}{dt}=\frac{\delta\mathcal{H}}{\delta \overline{a}_{\bm{k}}},
\end{align}
where $i=\sqrt{-1}$ and $\overline{a}_k$ is the complex conjugate of $a_k$. We use the notation $\delta\mathcal{G}/\delta \chi$ to refer to the first variation of the functional $\mathcal{G}$ with respect to the function $\chi$. The complex wave amplitude, $a_{\bm{k}}$, is the normal canonical transform of the Fourier transformed surface elevation, $\zeta_k=\mathcal{F}(\zeta)$, and the Fourier transformed velocity potential (evaluated at the surface), $\Psi_k=\mathcal{F}(\Psi)$. This canonical transform diagonalizes the Hamiltonian, reducing the formulation in two real variables to the above formulation in a single complex variable. The designation ``weak'' (or ``weakly nonlinear'') wave turbulence comes from the restriction of wave steepness, represented mathematically as $||\nabla\zeta||\ll1$~\cite{pushkarev_turbulence_2000}. The provision of this ``small parameter'' allows one to expand $\mathcal{H}$ in a perturbation series:
%
% hamiltonian expansion
\begin{align}
    \mathcal{H} = \mathcal{H}_2+\mathcal{H}_{\text{int}},
\end{align}
with the interaction Hamiltonian consisting of terms that are third or higher order in $a_{\bm{k}}$:
%
% hamiltonian expansion
\begin{align}
    \mathcal{H}_{\text{int}} = \mathcal{H}_3+\mathcal{H}_4+\mathcal{H}_5+...,
\end{align}
and representing the dynamical wave coupling that facilitates turbulent cascades of energy from small to large wavenumbers. 
            
The $\mathcal{H}_2$ term is 
\begin{align}
    \mathcal{H}_2=\int\omega_k\,a_{\bm{k}}\,\overline{a}_{\bm{k}}\,d\bm{k},
\end{align}
which represents freely traveling, non-interacting waves. This can be observed on substitution of $\mathcal{H}_2$ into eqn.~\eqref{eq:hamiltonian_equation}:
%
% non-interacting wave equation
\begin{align}
    i\,\frac{da_{\bm{k}}}{dt}=\omega_ka_{\bm{k}},
\end{align}
from which we observe that the $\mathcal{H}_2$ term describes linear processes. The general dispersion relation for surface tension dominated (\ie, capillary) water waves is
%
% general capillary waves dispersion relation
\begin{align}
    \omega_k^2=\frac{\sigma}{\rho}k^3\tanh \left(k h\right),
\end{align}
which satisfies the criteria: $\omega_0=0$, $\partial\omega_k/\partial k>0$, and $\partial^2\omega_k/\partial k^2>0$. This implies that capillary waves obey a decay type dispersion law and therefore correspond to a dominant $\mathcal{H}_3$ term in the expanded interaction Hamiltonian~\cite{zakharov_kolmogorov_1992}.

The amplitudes of \emph{all waves} encountered in this study are much smaller than the capillary length, $1/k_*$, so that surface tension effects are entirely dominant. Our interest is therefore only in those waves governed by the third-order interaction Hamiltonian:
%
% capillary wave hamiltonian
\begin{equation}\label{eq:third_order_hamiltonian}
    \mathcal{H}_3=\frac{1}{2}\int(V^{\bm{k}_1}_{\bm{k}_2,\bm{k}_3}\overline{a}_{\bm{k}_1}a_{\bm{k}_2}a_{\bm{k}_3}+\text{c.c.})\delta^{\bm{k}_1}_{\bm{k}_2+\bm{k}_3}\,d\bm{k}_1\,d\bm{k}_2\,d\bm{k}_3,
\end{equation}
where $V^{\bm{k}_1}_{\bm{k}_2,\bm{k}_3}=V^{\bm{k}_1}_{\bm{k}_3,\bm{k}_2}$ is an interaction coefficient with the noted symmetry property and ``c.c.'' denotes the complex conjugate of the preceding term. The expression $\delta^{\bm{b}}_{\bm{a}}$ may be interpreted as a Kronecker delta or as a Dirac delta with argument $\ltnorm{\bm{b}-\bm{a}}$. Then by substitution of the Hamiltonian into eqn.~\eqref{eq:hamiltonian_equation}, one has
\begin{align}\label{eq:reduced_hamiltonian_equation}
    i\,\frac{da_{\bm{k}}}{dt}-\omega_ka_{\bm{k}}=\frac{\delta\mathcal{H_{\text{int}}}}{\delta \overline{a}_{\bm{k}}}\approx\frac{\delta\mathcal{H}_3}{\delta \overline{a}_{\bm{k}}}.
\end{align}
After substituting eqn.~\eqref{eq:third_order_hamiltonian} into eqn.~\eqref{eq:reduced_hamiltonian_equation}, applying the variational derivative, and then simplifying, one obtains the three-wave equations of motion:
    \begin{widetext}
        \begin{equation}\label{eq:three_wave_eoms}
            i\,\frac{da_{\bm{k}}}{dt}-\omega_ka_{\bm{k}}=\int\left(\frac{1}{2}V^{\bm{k}}_{\bm{k}_1,\bm{k}_2}a_{\bm{k}_1}a_{\bm{k}_2}\,\delta^{\bm{k}}_{\bm{k}_1+\bm{k}_2}+\overline{V}^{\bm{k}_1}_{\bm{k},\bm{k}_2}a_{\bm{k}_1}\overline{a}_{\bm{k}_2}\,\delta^{\bm{k}_1}_{\bm{k}+\bm{k}_2}\right)\,d\bm{k}_1\,d\bm{k}_2.
        \end{equation}
    \end{widetext}

A fundamental result may be gleaned from eqns.~\eqref{eq:three_wave_eoms} by noting the Dirac combs, $\delta^{\bm{k}}_{\bm{k}_1+\bm{k}_2}$ and $\delta^{\bm{k}_1}_{\bm{k}+\bm{k}_2}$. When considered jointly, these ensure
\begin{align}
    \bm{k}_1\pm\bm{k}_2-\bm{k}_3=0,
\end{align}
where we have set $\bm{k}=\bm{k}_3$, following convention. In two dimensions, the capillary wave dispersion relation may be written as $\omega(\bm{k}_1)=\sqrt{(\sigma/\rho)\,(k_{1,x}^2+k_{1,y}^2)^{3/2}\tanh[(k_{1,x}^2+k_{1,y}^2)^{1/2}\,h]}$, where $\bm{k}_1=(k_{1,x},k_{1,y})^{\top}$ and $(\,\cdot\,)^{\top}$ is the transpose of $(\,\cdot\,)$. 

The decaying dispersion relation therefore defines a paraboloid, $P_1=\omega(\bm{k}_1)$, for variable $\bm{k}_1$. If one defines a second paraboloid centered on the $\bm{k}_2$ axis, $P_2=\omega(\bm{k}_2)$, then the intersection of these two surfaces, $P_1$ and $P_2$, in three dimensions satisfies the frequency relation $\omega(\bm{k}_1\pm\bm{k}_2)=\omega(\bm{k}_1)\pm\omega(\bm{k}_2)$ (\emph{see} Fig.~1.1 in Ref.~\cite{zakharov_kolmogorov_1992}). Then for fixed capillary wave vectors with $\omega_{k_n}=\omega(\bm{k}_n)$, one has the simultaneously satisfied relations
\begin{subequations}
    \begin{align}
        \bm{k}_1\pm\bm{k}_2-\bm{k}_3&=0,\\
        \omega_{k_1}\pm\omega_{k_2}-\omega_{k_3}&=0,
    \end{align}
\end{subequations}
known as the resonance conditions. These relations characterize the waves' merge ($2\rightarrow1$) or decay ($1\rightarrow2$) processes that define nonlinear capillary three-wave interactions.
            
To understand the overall behavior of the system, a statistical description is useful, and may be derived from the dynamics by considering the $p^{\text{th}}$ statistical correlator of the wave amplitude variable $a_{\bm{k}}$. The $p=1$ moment is $\langle a_{\bm{k}}\,\overline{a}_{\bm{k}'}\rangle=n(\bm{k})\,\delta^{\bm{k}}_{\bm{k}'}$, where one interprets $n(\bm{k})$ as the density of $k$ in k-space. When applied to eqn.~\eqref{eq:three_wave_eoms}, this operation leads to the three-wave kinetic equation
\begin{subequations}\label{eq:kinetic_equation}
    \begin{align}
        \frac{\partial\,n_{\bm{k}}}{\partial t}&=\mathcal{S}(n_{\bm{k}}),\\
        \mathcal{S}(n_{\bm{k}})&=R^{\bm{k}}_{\bm{k}_1,\bm{k}_2}+2R^{\bm{k_1}}_{\bm{k},\bm{k}_2},\\
        R^{\bm{k}}_{\bm{k}_1,\bm{k}_2}&=\pi\hspace{-0.1cm}\int\hspace{-0.1cm}|V^{\bm{k}}_{\bm{k}_1,\bm{k}_2}|f^{\bm{k}}_{\bm{k}_1,\bm{k}_2}\delta^{\bm{k}}_{\bm{k}_1+\bm{k}_2}\delta^{\omega_k}_{\omega_{k_1}+\omega_{k_2}}\,d\bm{k}_1\,d\bm{k}_2,\\
        f^{\bm{k}}_{\bm{k}_1,\bm{k}_2}&=n_{\bm{k}_1}n_{\bm{k}_2}-n_{\bm{k}}(n_{\bm{k}_1}+n_{\bm{k}_2}).
    \end{align}
\end{subequations}
For our purposes, the particular form taken by $V^{\bm{k}}_{\bm{k}_1,\bm{k}_2}$ depends on the depth of the medium. A well-known, fundamental result in the study of weak wave turbulence is the solution to the stationary three-wave kinetic equation, $\mathcal{S}(n_{\bm{k}})=0$, for capillary waves over deep water. Zakharov and Filonenko~\cite{zakharov_weak_1967} found the solution $n_{\bm{k}}\propto k^{-17/4}$.

As discussed in the introduction, at sufficiently large forcing inputs, NRB augments the resonance conditions to permit approximate satisfaction of the equations that model the interaction of the waves. At lower input powers, however, this broadening is relatively small. One may define a boundary between these regimes in terms of a discrete formulation of eqn.~\eqref{eq:three_wave_eoms}~\cite{lvov_discrete_2010}:
\begin{align}\label{eq:discrete_three_wave_eoms}
    &i\,\frac{da_{\bm{k}}}{dt}-\omega_ka_{\bm{k}}=\notag\\
    \hspace{-0.15cm}\sum_{\bm{k}_1,\bm{k}_2}&\hspace{-0.1cm}\left(\frac{1}{2}V^{\bm{k}}_{\bm{k}_1,\bm{k}_2}a_{\bm{k}_1}a_{\bm{k}_2}R^{\bm{k}}_{\bm{k}_1+\bm{k}_2}+\overline{V}^{\bm{k}_1}_{\bm{k},\bm{k}_2}a_{\bm{k}_1}\overline{a}_{\bm{k}_2}R^{\bm{k}_1}_{\bm{k}+\bm{k}_2}\right),
\end{align}
where $R^{\bm{k}}_{\bm{k}_1+\bm{k}_2}$ evaluates to unity under satisfaction of the resonance conditions in the absence of NRB.

% FIGURE TWO
% TWO COLUMN FIGURE
% power spectral density classification plots
\begin{figure*}[htb]
    \begin{center}
        \includegraphics[width=\textwidth]{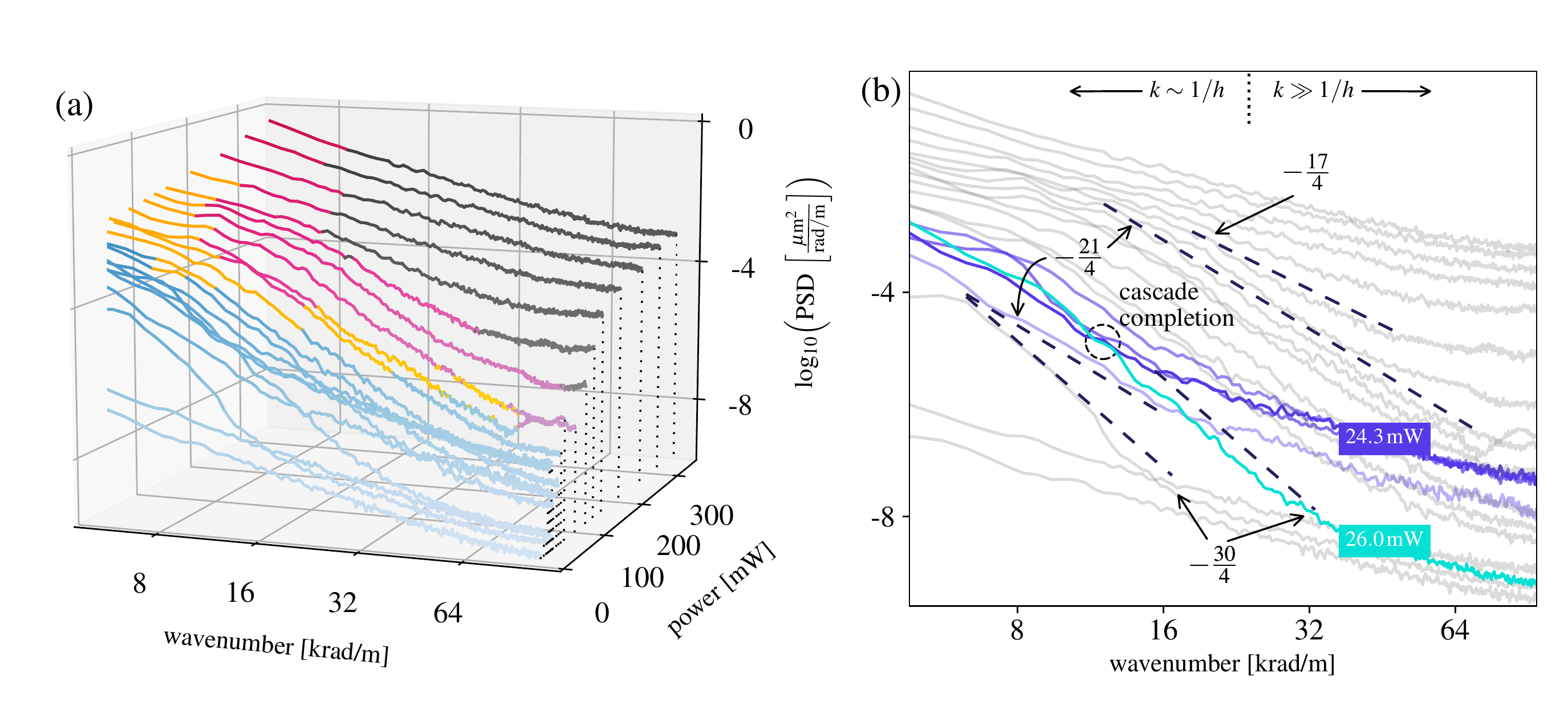}
        \caption{Turbulent micro capillary wave power spectral density regime classification. (a) The measured power spectral density, plotted with respect to wavenumber as functions of increasing power may be classified in terms of their WT regimes: (blue) discrete, (yellow) intermediate,  (pink) kinetic, and (gray) strong. The first three are defined in eqns.~\eqref{eq:shallow_regimes} and \eqref{eq:deep_regimes}. The fourth is discussed later and demonstrated in Fig.~\ref{fig:f03_regime_map}. (b) An input of $24.3\,$mW leads to an arrested cascade. Increasing the power to $26\,$mW leads to cascade completion, demonstrating a critical NRB value as theorized in Ref.~\cite{connaughton_discreteness_2001}. Also shown (light to dark purple lines) are three input powers preceding $24.3\,$mW: $15.5\,$mW, $18.5\,$mW, and $21.5\,$mW. The indicated slopes are $17/4$, the kinetic capillary wave slope predicted by Zakharov~\cite{zakharov_weak_1967}; the steepest slope, $21/4$, observed in simulations of Ref.~\cite{pan_understanding_2017} and corresponding to an approximately constant slope over the KWT regime here (as discussed later); and the steepest slope, $30/4$, observed in our experiments, both at capillary wave onset and immediately after the critical NRB value for a $26\,$mW  input. Spectra in these plots are generated using Welch's method with Hann windowing and fifty percent overlap to average roughly one hundred spectra for all but the lowest two powers ($0$ and $7\,$mW). The shallow regime (not shown) exists at wavenumbers $\lesssim2\,$krad/m.}
        \label{fig:f02_spectra}
    \end{center}
\end{figure*}

%--------------------------------------------------
% INTRA-SPECTRUM REGIME CLASSIFICATION
%--------------------------------------------------
\subsection{Intra-spectrum regime classification}
Using the WWT theory, we next derive and apply dimensionless parameters to quantitatively classify WT regimes observed in our surface measurements. These parameters also identify the appropriate modeling approach to later use in analysis of the wave phenomena. When DWT dominates, resonance conditions are Diophantine, and number theoretical approaches are necessary~\cite{kartashova_nonlinear_2010}. If KWT dominates, resonances are approximate and the kinetic theory is valid~\cite{kartashova_discrete_2009}.
            
The appropriate formulation of the NRB depends on the wave turbulent regime being analyzed, with  NRB in KWT expressed via the kinetic equation \eqref{eq:kinetic_equation} and NRB in DWT expressed via the discrete equations of motion \eqref{eq:discrete_three_wave_eoms}. The NRB parameter is given for these two cases by~\cite{lvov_discrete_2010}
%
% discrete nonlinear resonance broadening
\begin{align}\label{eq:nrb}
    \delta_{\textsc{NRB}} =
    \begin{cases}
        |V_k\,a_k|\,\mathcal{N}_k,&\quad(\text{DWT})\\
        |V_k|^2|a_k|^2(k\,L)^2/\omega_k, &\quad(\text{KWT})
    \end{cases}
\end{align}
where $|(\,\cdot\,)|$ denotes the complex modulus. The parameter $V_k$ is a simplified approximation of the interaction coefficient; it only depends upon the wavenumber. The canonical amplitude variable $a_k$ is related to the orthonormal Fourier amplitude~\cite{pushkarev_turbulence_2000}:
%
% canonical transform
\begin{align}\label{eq:canonical_transform}
    \zeta_k = \sqrt{\frac{\rho\,\omega_k}{2\,\sigma\,k^2}}(a_k+a^*_{-k}),
\end{align}
where $\sigma$ is the interfacial surface tension and $\rho$ is the fluid density. Note that eqn.~\eqref{eq:canonical_transform} assumes normalization of the Fourier transform by $\sqrt{2\,\pi}$. Here $\mathcal{N}_k$ is the number of exact resonances that are dynamically relevant. For the capillary wave systems we consider here, wavenumber locality is assumed---$k_1\sim k_2\sim k_3$---so that $\mathcal{N}_k\gtrsim1$. We set $\mathcal{N}_k=1$ for our order-of-magnitude analysis in order to avoid unnecessary complexity.
        
The particular forms of the dispersion relation and the interaction coefficient depend on the depth of the fluid. They determine the form our classification parameters ultimately take for a given data set. Our system progresses from shallow at $\mathcal{O}(10^2)\,$rad/m to deep at $\mathcal{O}(10^4)\,$rad/m. Intermediate regimes are defined by interpolating between the shallow and deep conditions (\emph{see} [supplemental materials]).

%--------------------------------------------------
% SHALLOW WATER CLASSIFICATION
%--------------------------------------------------
\subsubsection{Shallow water classification ($k_*\ll k \ll 1/h$)}
The shallow water capillary wave dispersion relation is
%
% shallow water dispersion
\begin{align}\label{eq:shallow_dispersion}
    \omega_k = \sqrt{\frac{\sigma\,h}{\rho}}k^2.
\end{align}
The interaction coefficient for shallow water waves is~\cite{zakharov_kolmogorov_1992}
%
% shallow interaction coefficient
\begin{align}\label{eq:shallow_coeff}
    V_k = \frac{k^2}{8\,\pi}\left(\frac{\sigma}{4\,\rho\,h}\right)^{1/4}.
\end{align}
Combining eqns.~\eqref{eq:grid_spacing}, \eqref{eq:nrb}--\eqref{eq:shallow_coeff}, along with the respective requirements on $\delta_{\textsc{NRB}}$ and $\delta_{k}$, we produce the following shallow-water WT regimes:
%
% shallow water WT regimes
\begin{subequations}\label{eq:shallow_regimes}
    \begin{align}
        &\text{DWT}\quad\text{if}\quad\Delta_s = \tfrac{1}{16\,\pi}\mathcal{A}_v/\mathcal{A}_w \ll 1,\\
        &\text{KWT}\quad\text{if}\quad\Lambda_s = 2\,\pi^2\,\mathcal{S}_w\,\Delta_s^2 \gg 1,\label{eq:shallow_kinetic}\\
        &\text{IWT}\quad\text{otherwise},
    \end{align}
\end{subequations}
where $\mathcal{A}_w=\frac{1}{k\,\widehat{\zeta}_k}$%1/k\,\widehat{\zeta}_k$  --- ambiguous solidus
is the wave aspect ratio, $\mathcal{A}_v=\frac{1}{k_m\,h}$%1/k_m\,h$ --- ambiguous solidus
is the quiescent fluid volume aspect ratio, and $\widehat{\zeta}_k$ is the Fourier-transformed wave amplitude. Referring to Fig.~\ref{fig:f01_exp_setup}, the grid spacing is $k_m=\pi/D_{\textrm{inner}}$. We define the wave seclusion as $\mathcal{S}_w=k/k_m$, with $\mathcal{S}_w\gg1$ indicating negligible boundary effects. 

Essentially, the condition for shallow water DWT is that the quiescent fluid should be much deeper than the waves. For KWT, the waves should be deep in comparison to the quiescent fluid's depth, and the waves should be isolated from boundary effects.

%--------------------------------------------------
% DEEP WATER CLASSIFICATION
%--------------------------------------------------
\subsubsection{Deep water classification ($k \gg 1/h$)}

The deep water capillary wave dispersion relation is
%
% deep water dispersion
\begin{align}\label{eq:deep_dispersion}
    \omega_k^2 = \frac{\sigma}{\rho}k^3.
\end{align}
By pairing the assumption of wavenumber locality with an order-of-magnitude analysis, we approximate the interaction coefficient~\cite{pushkarev_turbulence_2000} for capillary waves in deep water:
%
% deep interaction coefficient
\begin{align}\label{eq:deep_coeff}
    V_k \approx \frac{1}{8\,\pi}\sqrt{\frac{\rho\,\omega_k^3}{2\,\sigma}}.
\end{align}
Combining eqns.~\eqref{eq:grid_spacing}, \eqref{eq:nrb}--\eqref{eq:canonical_transform}, and \eqref{eq:deep_dispersion}--\eqref{eq:deep_coeff} with the respective requirements on $\delta_{\textsc{NRB}}$ and $\delta_{k}$, we obtain the following deep-water WT regimes:
%
% deep water WT regimes
\begin{subequations}\label{eq:deep_regimes}
    \begin{align}
        &\text{DWT}\quad\text{if}\quad\Delta_d = \tfrac{1}{12\,\pi}\mathcal{S}_w/\mathcal{A}_w \ll 1,\\
        &\text{KWT}\quad\text{if}\quad\Lambda_d = \tfrac{3\,\pi^2}{2}\mathcal{S}_w\,\Delta_d^2 \gg 1,\label{eq:deep_kinetic}\\
        &\text{IWT}\quad\text{otherwise}.
    \end{align}
\end{subequations}
Thus, in deep water only the lateral dimension is relevant, whereas in shallow water (eqn.~\eqref{eq:shallow_regimes}) the definition of the regimes depend on both dimensions. For deep water, finite-basin effects occur when the wave steepness is large relative to the domain's breadth. Moreover, in deep water, the KWT condition has weaker wave seclusion requirements and stronger wave steepness requirements. This qualitatively agrees with expressions found elsewhere in the literature~\cite{nazarenko_sandpile_2006,denissenko_gravity_2007}. Equations~\eqref{eq:shallow_regimes}~and~\eqref{eq:deep_regimes} and the wave turbulence regimes they define depend upon the wavenumber. Fundamentally, the wave turbulence regimes are \emph{geometrically determined}.

%--------------------------------------------------
% INTRA-SPECTRUM RESULTS ANALYSIS
%--------------------------------------------------
\subsubsection{Intra-spectrum results analysis}
It is possible to demarcate capillary wave turbulence regimes based upon changes in the power law representations between the regimes. Figure~\ref{fig:f02_spectra} outlines the spectral features of our system written in terms of power spectra, $S$. In shallow regions, $S_{\textsc{D-I}}\propto k^{-2}$ and $S_{\textsc{I-K}}\propto k^{-3}$, respectively, for the DWT-IWT (D-I) and the IWT-KWT (I-K) PSD bounds. In deep regions, $S_{\textsc{D-I}}\propto k^{-4}$ and $S_{\textsc{I-K}}\propto k^{-5}$. The change is a consequence of the continuously increasing effect of $\delta_{\textsc{NRB}}$ within the finite-depth basin as energy is redistributed to larger wavenumbers in the cascade. The change in the regime bound definitions---\ie, proxies for the character of the WT---will smoothly and monotonically vary between these regimes.

Detailed inspection of these spectra reveals order-of-magnitude agreement of regime bounds with expected spectral characteristics. Theoretically, the hallmark of DWT is arrested cascades that are incrementally extended by NRB through sandpiling~\cite{nazarenko_sandpile_2006}. The irregular lumpy appearance visible at lower input powers ($P\lesssim24\,$mW) in Fig.~\ref{fig:f02_spectra} indicates energy confinement due to cascade arrest: ``frozen turbulence''. This regime is defined primarily by spectral volatility. As the power is increased towards the DWT-IWT boundary, the  shape and amplitude of the spectra remain approximately constant until NRB overcomes the discrete modal separation, as theorized by Connaughton \etal~\cite{connaughton_discreteness_2001}. Beyond this threshold power defined by the characteristics of the system---$P\gtrsim26\,$mW here---energy flows freely to small scales, leading to complete cascades, \ie, cascades that extend to the viscous dissipation range near the noise floor.

The IWT regime is brief and characterized by residual irregularities imposed by discreteness and a more easily discernible slope due to KWT onset. At still higher powers, spectral fluctuations are mostly abated and the spectrum essentially has a continuous appearance associated with KWT. In this regime, we expect Zakharov's kinetic theory~\cite{zakharov_weak_1967} to hold to good approximation. Although one finds evidence of a constant spectral slope, the slope value---approximately $-21/4$---is different than the $-17/4$ law theorized by Zakharov. This is true despite the otherwise good agreement between the data and the characteristics of the kinetic theory in this regime. At even higher powers, the spectra depart from the constant $-21/4$ slope, signaling a change in regime from weak KWT to SWT. It is evident in Fig.~\hyperref[{fig:f02_spectra}]{2(a)} that SWT initially appears at higher wavenumbers and spreads toward lower wavenumbers as the power is increased.

The steepest spectral slope, $-30/4$, occurs twice over the range of investigated powers. Its first appearance is at the onset of continuous capillary wave motion. This is just beyond a range of input powers where the dynamics are intermittent, punctuated by brief quiescent periods. As the power increases beyond this onset value, the cascade remains truncated but its slope gradually increases to a maximum value that persists as the power is further increased to 24.3~mW. With a small subsequent increase in input power ($\sim1.5\,$mW) to 26.0~mW, the cascade abruptly completes, reinstating the initial slope of $-30/4$. If the power is further increased, the slope gradually increases to $\approx-21/4$ and persists at this value up to an input power of $125\,$mW. Beyond $125\,$mW, the spectral slope again increases with an increase in power, approaching an apparently constant value of $\approx-2$ in the SWT regime. This feature, which the authors are unable to compare with other studies from the literature, suggests a power-law type solution within a subset of the SWT regime. We note that the foregoing results differ from those obtained in simulations~\cite{pan_understanding_2017}. The slopes in the latter range from roughly $-21/4$ at lower powers to $-17/4$ at the highest powers and were obtained using Zakharov's kinetic equations modified to account for the discreteness of the domain.

%--------------------------------------------------
% GLOBAL NONLINEARITY ANALYSIS 
%--------------------------------------------------
\subsection{Global nonlinearity analysis}
Over the years, attempts have been made at defining an overall measure of the nonlinearity of WT. With the benefit of an extensive data set, we also attempt to define a global measure that quantifies WT nonlinearity with a single value:
%
% delta-mu measure
\begin{align}\label{eq:delta_mu}
    \delta_{\mu}=\sqrt{\frac{\sum_{n=0}^N\sum_{m=0}^n\overline{\delta}_n\,\overline{\delta}_m\,b_{n,m}^2}{\sum_{n=0}^N\sum_{m=0}^nb_{n,m}^2}}.
\end{align}
Here $\overline{\delta}_n=\delta_{\textsc{NRB}}(f_n)/\delta_{k_n}$ quantifies nonlinearity at wavenumber $k$ using the ratio of resonance broadening to grid spacing. The root-mean-square summation in eqn.~\eqref{eq:delta_mu} is weighted by the bicoherence. Since bicoherence is defined over a symmetric frequency grid, we take double summation over unique pairings. The bicoherence is
%
% bicoherence
\begin{align}\label{eq:bicoherence}
    b_{n,m}=\frac{|\langle\widetilde{\zeta}(t,f_n)\,\widetilde{\zeta}(t,f_m)\,\widetilde{\zeta}^*(t,f_n+f_m)\rangle_t|}{\langle|\widetilde{\zeta}(t,f_n)\,\widetilde{\zeta}(t,f_m)\,\widetilde{\zeta}^*(t,f_n+f_m)|\rangle_t}.
\end{align}
The sole difference between the numerator and denominator in this definition is in the order of operations between the time average $\langle\,\cdot\,\rangle_t$ and the modulus $|\,\cdot\,|$. The numerator is the bispectrum. In the bispectrum, phase information is removed after time averaging, such that this describes the average wave coupling strength at a particular point in frequency space. In the denominator of eqn.~\eqref{eq:bicoherence}, the phases are set to zero (\ie, made equivalent) before time averaging, representing perfect phase coupling. Thus, bicoherence, $b\in[0,1]$, measures the \emph{normalized} three-wave coupling strength, and is relevant for analysis of capillary waves specifically. Equation~\eqref{eq:delta_mu} is similar to a metric used by Pan and Yue~\cite{pan_understanding_2017}. However, we directly operate on all three frequencies (in a manner similar to Ref.~\cite{aubourg_investigation_2016}) instead of Pan and Yue's approach of fixing a wavevector within the three-wave dependence~\cite{pan_understanding_2017}. Since bicoherence obtained from Fourier methods is noisy and insufficient for use with turbulence spectra, we follow van Milligen~\etal~\cite{van_milligen_nonlinear_1995} in using wavelet-based transforms. The wavelet transforms $\zeta(t)\mapsto\widetilde{\zeta}(t,f)$ applied to the surface displacement measurements provides the instantaneous spectra needed to compute eqn.~\eqref{eq:bicoherence}. Implementation details are provided in the [supplemental materials]. 
 
% FIGURE THREE
% ONE COLUMN FIGURE
% nonlinearity-based regime map
\begin{figure}
    \begin{center}
        \includegraphics[width=\columnwidth]{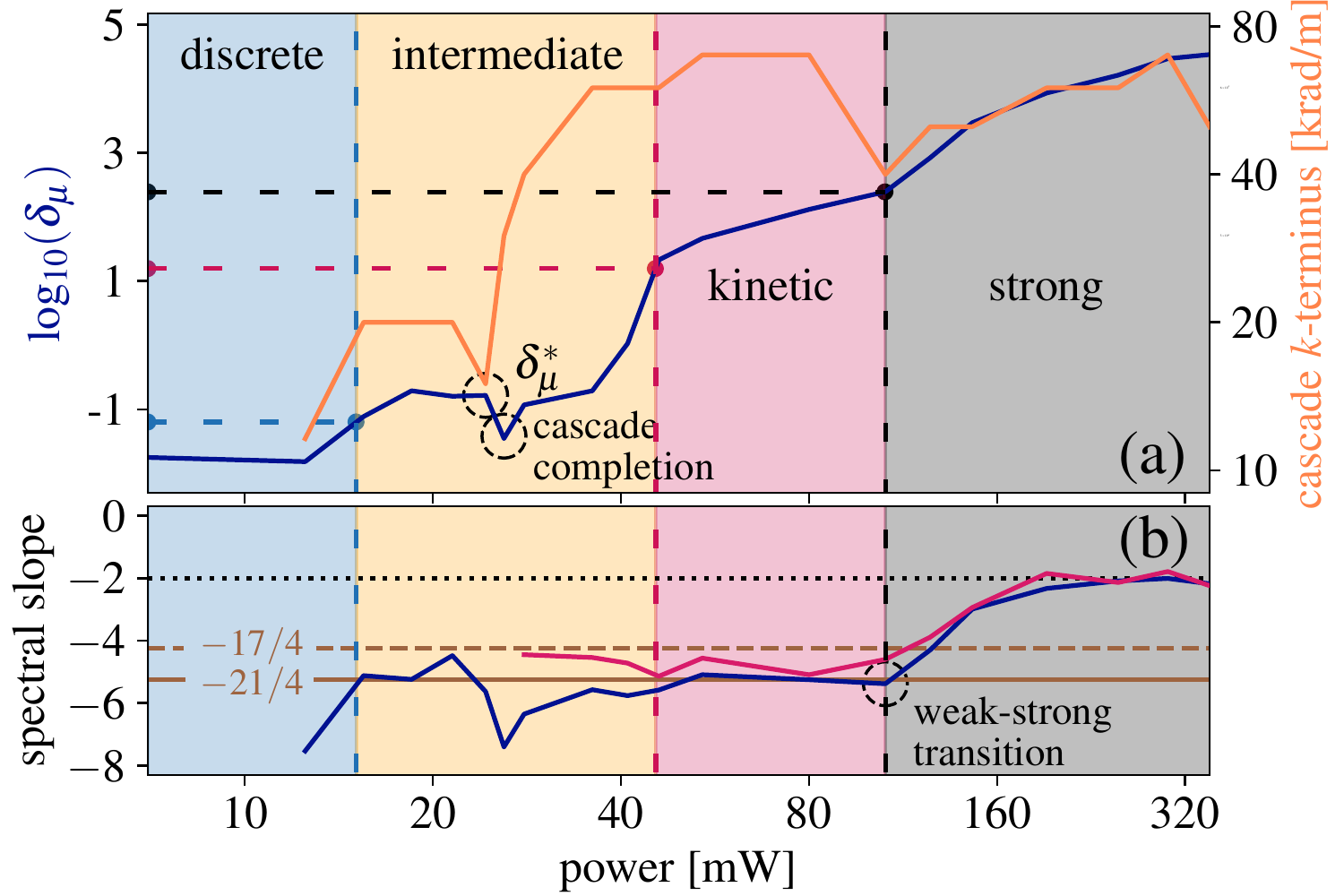}
        \caption{Global nonlinearity map of variable-regime capillary WT. (a) Nonlinearity level eqn.~\eqref{eq:delta_mu} and terminal cascade wavenumber as functions of increasing input power. (b) Inertial regime spectral slope over (blue line) the entire cascade and (red line) the deep water regime. As the power continues to increase, a region of approximate constancy in nonlinearity, cascade length, and slope is observed just before the critical value, $\delta^*_{\mu}\approx0.2$, beyond which a decrease of all three parameters occurs. A sharp increase in cascade length immediately follows $\delta^*_{\mu}\approx0.2$. The KWT-SWT bound occurs when the spectral slope exceeds the experimentally constant value, $-21/4$, as observed within the kinetic regime.}
        \label{fig:f03_regime_map}
    \end{center}
\end{figure}
With eqn.~\eqref{eq:delta_mu} as our definition for the nonlinearity parameter, $\delta_{\mu}$, we find a non-monotonic structure that may be correlated to key features of the WT as provided in Fig.~\ref{fig:f03_regime_map}. The terminal cascade wavenumber and the spectral slope are plotted along with the nonlinearity parameter versus the input power. Since Zakharov's kinetic theory is based on assumptions of deep water waves and weak nonlinearity, the transition to strong nonlinearity is evident when the observed constant kinetic slope $\gamma$ exceeds $21/4$, and corresponding to a nonlinearity parameter value of $\mathcal{O}(\delta_{\mu})=10^2$. This suggests a boundary between weak and strong nonlinearity as follows:
%
% strong wave turbulence bound
\begin{align}
    \text{SWT}\quad\text{if}\quad\Lambda/10 \gg 1,
\end{align}
where $\Lambda$ is obtained from eqn.~\eqref{eq:shallow_kinetic}, eqn.~\eqref{eq:deep_kinetic}, or the interpolated region between. Using this criteria, strongly nonlinear wave spectra have been identified in Fig.~\hyperref[{fig:f02_spectra}]{1(b)} in gray, appearing first at large wavenumbers, and spreading toward smaller wavenumbers as input power is increased.
    
            Figure~\ref{fig:f03_regime_map} reveals many significant features beyond those already mentioned. At a $16\,$mW input---corresponding to the DWT-IWT bound---a region of approximately constant nonlinearity level, cascade length, and slope value exists that corresponds to the frozen turbulence region. This immediately precedes a critical value, $\delta^*_{\mu}\approx0.2$, that causes cascade completion, as indicated by a very sharp increase in the terminal cascade wavenumber and an abrupt decrease in both $\delta_{\mu}$ and slope. The entirety of the critical transitional process occurs within the IWT regime, during which the nonlinearity level ``stalls,'' remaining approximately constant at $\delta_{\mu}\ll1$.
    
The pattern of build-up, plateau, and decrease of the cascade $k$-terminus is initially repeated on entry to the KWT regime. However, the system enters into the SWT regime without  significant decreases in the slope or nonlinearity level nor a sharp increase in the cascade's terminal wavenumber. The terminal wavenumber in and beyond the transition region is $50-70\,$krad/m, so the associated Kolmogorov scales are $\lambda\sim100\,\upmu$m, $\varepsilon\sim0.1\,$mW/$\upmu$L, $\tau\sim100\,\upmu$s, and $u\sim100\,$cm/s. On transition into the SWT regime, the changing slope resembles the pattern of slope change beginning at $\delta^*_{\mu}$ and attains a new plateau of $-2$. This value of the slope may assume the same role in the SWT regime that the Zakharov slope assumes in the KWT regime.

%--------------------------------------------------
% CONCLUSIONS
%--------------------------------------------------
\subsection{Conclusions}
Our analysis has shown that a thinly-wetted surface driven by HFUS is a dynamically rich system. The dynamics traverse several WT regimes and, at sufficient powers, reach the strongly nonlinear regime. In atomizing systems, aerosol dispersal occurs at levels of nonlinearity that are inaccessible to modern WT theories, let alone classic interpretations of atomization based on weakly turbulent phenomena. Further study of strongly nonlinear WT is therefore essential to a rigorous understanding of the atomization phenomenon.

%--------------------------------------------------
% ACKNOWLEDGEMENTS
%--------------------------------------------------
\section*{Acknowledgements}
We are grateful to the Office of Naval Research (grant 12368098) and the W.M.\ Keck Foundation for funding provided to J.\ Friend in support of this work. J.\ Orosco is grateful for support provided by the University of California's Presidential Postdoctoral Fellowship program. We are also grateful to Yves Emery and team at Lynce\'e~Tec for assistance with adapting the DHM to this project's needs. Fabrication was performed in part at the San Diego Nanotechnology Infrastructure (SDNI) of UCSD, a member of the National Nanotechnology Coordinated Infrastructure, which is supported by the National Science Foundation (Grant ECCS–1542148).

\newpage

%--------------------------------------------------
% BIBFILE
%--------------------------------------------------
% \bibliographystyle{unsrtnat}
% \bibliographystyle{plain}
\bibliographystyle{elsarticle-num}
\bibliography{wtid}

% end of the doc
\end{document}